\documentclass[12pt,aps,prd,showpacs,amsmath,amssymb]{revtex4}
\input epsf
\begin{document}
\title{Light-cone QCD Sum Rules for the $\Lambda$ Baryon Electromagnetic Form Factors and its magnetic moment}
\author{Yong-Lu Liu and Ming-Qiu Huang}
\affiliation{Department of Physics, National University of Defense Technology, Hunan 410073, China}
\date{\today}
\begin{abstract}
We present the light-cone QCD sum rules up to twist $6$ for the
electromagnetic form factors of the $\Lambda$ baryon. To estimate
the magnetic moment of the baryon, the magnetic form factor is
fitted by the dipole formula. The numerical value of our estimation
is $\mu_\Lambda=-(0.64\pm0.04)\mu_N$, which is in accordance with
the experimental data and the existing theoretical results. We find
that it is twist $4$ but not the leading twist distribution
amplitudes that dominate the results.
\end{abstract}
\maketitle

\section{Introduction}
\label{sec1}Electromagnetic (EM) form factors of the hadron are
fundamental objects, for they characterize the internal structure of
the composite particle. There were a lot of experimental results on
EM form factors of baryons \cite{Walker,Arrington,Bosted,Lung} and
mesons \cite{Jebek,Dally,Liesenfeld,Volmer,Horn,Tadevosyan} in the
past decades. Due to the complexity of their structure, theoretical
investigations on baryons received much less attention than those on
mesons, and the existing studies on EM form factors of baryons were
mainly focused on the nucleon. In comparison with the nucleon, so
far as we know, there are no experimental data for $\Lambda$ at
present. Therefore, it is instructive and necessary to study its EM
form factors theoretically. Fortunately, there have been
experimental result on the magnetic moment and other model-dependent
theoretical results on EM form factors to test our estimation
indirectly. Van Cauteren $et$ $al.$ have investigated the electric
and magnetic form factors of strange baryons in the relativistic
constituent-quark model \cite{Cauteren}. The chiral perturbation
theory \cite{Kubis} and the chiral quark/soliton model \cite{Kim}
have been used to investigate the $\Lambda$ EM form factors at low
momentum transfer. The present work is devoted to investigate the
$\Lambda$ EM form factors at moderately large momentum transfer and
estimate its magnetic moment theoretically.

The matrix element of the EM current between the initial and final
$\Lambda$ baryon states can be parameterized in terms of the Dirac
and Pauli form factors $F_1(Q^2)$ and $F_2(Q^2)$:
\begin{eqnarray}
\langle\Lambda(P')|j_\mu^{em}(0)|\Lambda(P)\rangle=\bar \Lambda
(P')[\gamma_\mu F_1(Q^2) -i\frac{\sigma_{\mu\nu}
q^\nu}{2M_\Lambda}F_2(Q^2)]\Lambda(P),\label{form}
\end{eqnarray}
where $j_\mu^{em}=e_u\bar u\gamma_\mu u+e_d \bar d\gamma_\mu d+e_s
\bar s \gamma_\mu s$ is the EM current relevant to the baryon. $P$
and $P'$ are the four-momenta of the initial and final $\Lambda$
baryon states, $M_\Lambda$ is the $\Lambda$ mass,
$P^2={P'}^2=M_\Lambda^2$, $Q^2=-q^2=-(P-P')^2$ is the momentum
transfer (outgoing photon momentum), and $\Lambda(P)$ is the
$\Lambda$ spinor. From the experimental point of view, Dirac and
Pauli form factors are described equivalently by another set of form
factors: electric $G_E(Q^2)$ and magnetic $G_M(Q^2)$ Sachs form
factors:
\begin{eqnarray}
G_E(Q^2)&=&F_1(Q^2)-\frac{Q^2}{4M^2}F_2(Q^2), \nonumber\\
G_M(Q^2)&=&F_1(Q^2)+F_2(Q^2).
\end{eqnarray}
The normalization of the Sachs form factors at $Q^2=0$ is given by
the baryon charge $G_E(0)=0$ and magnetic moment
$G_M(0)=\mu_\Lambda$. In the Breit frame, $G_E(Q^2)$ corresponds to
the distribution of the electric charge and $G_M(Q^2)$ to the
magnetic current distribution.

To give the EM form factors, QCD sum rule method from a three-point
correlation function is a useful tool. However, the three-point sum
rule has a deficiency which restricts its application and accuracy
\cite{lcsr}. Instead of it, the light-cone sum rule (LCSR) method,
which has firstly been employed on the nucleon by Braun $et$ $al.$
\cite{Lenz}, is utilized to study the $\Lambda$ EM form factors in
this paper. LCSR, a hybrid of the traditional sum rule (SVZ sum
rule) \cite{SVZ} and the theory of hard exclusive processes, is a
useful method to include both hard scattering and soft contributions
in QCD. It was developed in late 1980's by making a partial
resummation of the operator product expansion (OPE) to all orders
\cite{lcsr1,lcsr2,lcsr3}. The main difference between SVZ sum rule
and LCSR is that the short-distance Wilson OPE in increasing
dimension is replaced by the light-cone expansion in terms of
distribution amplitudes (see Ref. \cite{Colangelo} for a review) of
increasing twist. The $\Lambda$ baryon distribution amplitudes,
which are the fundamental input parameters in LCSR, have been given
in Ref. \cite{Chernyak,Wang}. In this paper, we give a little
correction to the nonperturbative parameter $\lambda_1$ which has
different sign compared with that in Ref. \cite{Wang}. To leading
order in the QCD coupling $\alpha_s$, we give the $Q^2$-dependence
EM form factors of $\Lambda$ in the range $1\;\mbox{GeV}^2\le
Q^2\le7\;\mbox{GeV}^2$ considering contributions up to twist $6$.

Another aim of this manuscript is to estimate the $\Lambda$ magnetic
moment. As important intrinsic physical values, the magnetic moments
of baryons have been investigated theoretically in the past years by
various models
\cite{Kubis,Pasupathy,Franklin,Kerbikov,Puglia,Park,Contreras,Simonov,Bartelski,Aliev,Aliev1,Aliev2}.
Among these models, QCD sum rule is widely used. The conventional
way to calculate the magnetic moments of hadrons in QCD sum rule is
to deal with the two-point correlation function in the background of
the electromagnetic field at zero momentum transfer. This is the
developed QCD sum rule method proposed by Balitsky $et$ $al.$
\cite{Balitsky} and Ioffe $et$ $al.$ \cite{Ioffe}. In their method,
the correlation function is expanded in a constant weak external
electromagnetic field $F_{\mu\nu}$, and the sum rule of the term
connected with the magnetic moment can be gotten directly. This
method has been widely used in calculating the magnetic moments of
hadrons
\cite{Pasupathy,Balitsky,Ioffe,Wilson,Lee,Zhu,Samsonov,Chiu}. In the
present paper, another approach is adopted to evaluate the $\Lambda$
magnetic moment. The basic idea of the approach is to extract the
magnetic moment from the calculated EM form factors indirectly. It
is assumed that the magnetic form factor divided by the magnetic
moment can be described by the dipole formula:
\begin{equation}
\frac{1}{\mu_\Lambda}G_M(Q^2)=\frac{1}{(1+Q^2/m_0^2)^2}=G_D(Q^2).\label{dipole}
\end{equation}
Since there are no experimental data available for the form factor,
the result is fitted by the above dipole formula, and the two
parameters $\mu_\Lambda$ and $m_0^2$ are artificial parameters to be
determined simultaneously. After that, we use the fit to estimate
the magnetic moment of the baryon. In Ref. \cite{Aliev1}, the
authors have ever calculated the magnetic moment of the $\Lambda$
baryon with LCSR. In their calculations, the magnetic moment can be
obtained directly. This comes from the fact that they use the photon
distribution amplitudes, which allows for the utilization of LCSR at
the point $Q^2=0$. However, in our case, the working region of the
sum rule cannot extrapolate to the zero point directly. Hence, we
have to estimate it with other approaches.

This paper is organized as follows. Section \ref{sec2} is devoted to
derive the light-cone QCD sum rules of the form factors relevant to
the momentum transfer $Q^2$, and the necessary distribution
amplitudes needed in the calculation are presented in this section.
Section \ref{sec3} is the numerical analysis part in which the
magnetic form factor is fitted by the dipole formula, and the
magnetic moment is estimated from the fit. This section also
presents the QCD sum rules of the nonperturbative parameters
$f_\Lambda$ and $\lambda_1$. The summary and conclusion are given at
the end of this section.
\section{ Light-cone QCD Sum Rules for the EM form factors}
\label{sec2} \subsection{Leading twist} The derivation of the sum
rules begins with the following correlation function:
\begin{equation}
T_\mu(P,q)=i \int d^4xe^{iq\cdot x}\langle
0|T\{j_\Lambda(0)j_\mu^{em}(x)\}|\Lambda(P)\rangle,\label{correlator}
\end{equation}
where the interpolating current of the $\Lambda$ baryon field is
chosen to be
\begin{equation}
{j_\Lambda}(0)=\epsilon_{ijk}(u^i(0)C\gamma_5\!\not\!
zd^j(0))\!\not\! zs^k(0).
\end{equation}
The coupling constant is defined by the matrix element of the
interpolating current between the vacuum and the $\Lambda$ state:
\begin{equation}
\langle 0|j_\Lambda|\Lambda\rangle=f_\Lambda(P\cdot z)\!\not\!
z\Lambda(P),\label{norm}
\end{equation}
where $z$ is a light-cone vector, $z^2=0$, and $f_\Lambda$
determines the normalization of the leading twist $\Lambda$
distribution amplitude.

In order to get the sum rules, the correlation function
(\ref{correlator}) needs to be expressed both phenomenologically and
theoretically. On one side, by inserting a complete set of
intermediate states with the same quantum numbers as those of
$\Lambda$, and using the definition of the form factors in Eq.
(\ref{form}) and the definition in Eq. (\ref{norm}), the hadronic
representation of the correlation function is expressed as follows:
\begin{eqnarray}
z^\mu T_\mu(P,q)&=&\frac{1}{M_\Lambda^2-P'^2}f_\Lambda (P'\cdot
z)[2(P'\cdot zF_1(Q^2)\nonumber\\
&&-\frac{q\cdot z}{2}F_2(Q^2))\!\not\! z+(P'\cdot
zF_2(Q^2)\nonumber\\&& +\frac{q\cdot z}{2}F_2(Q^2))\frac{\!\not\!
z\!\not\! q}{M_\Lambda}]\Lambda(P)+...\;,
\end{eqnarray}
where $P'=P-q$, and the dots stand for the higher resonance and
continuum contributions. Here the contraction of the correlation
function by $z^\mu$ is adopted to get rid of contributions which are
subdominant on the light cone.

On the other side, the correlation function is calculated in
perturbation theory at large Euclidean momenta $P'^2$ and $q^2=-Q^2$
in terms of the distribution amplitudes. To the leading order, the
theoretical side of the correlation function is
\begin{eqnarray}
z^\mu T_\mu(P,q)&=&2(P\cdot z)^2\!\not\! z\Lambda(P)\int
\mathcal{D}\alpha
\{e_u\frac{\mathcal{A}_1(\alpha)}{s_1-(q-P)^2}\nonumber\\
&&+e_d\frac{\mathcal{A}_1(\alpha)}
{s_2-(q-P)^2}+e_s\frac{\mathcal{A}_1(\alpha)}{s_3-(q-P)^2}\},\nonumber\\
\end{eqnarray}
where $s_i=(1-\alpha_i)P^2-(1-\alpha_i)/\alpha_iq^2+m_i^2/\alpha_i$,
and $m_{1,2}=0,m_3=m_s$. The $\Lambda$ baryon distribution
amplitudes are defined by the matrix element of the operator between
vacuum and the $\Lambda$ state. The leading order distribution
amplitudes are expressed as
\begin{eqnarray}
&&4\langle0|u_\alpha^i(a_1x)d_\beta^j(a_2x)s_\gamma^k(a_3x)|\Lambda(P)\rangle\nonumber\\
&=&\mathcal{V}_1(\!\not\! P
C)_{\alpha\beta}(\gamma_5\Lambda)_\gamma+\mathcal{A}_1(\!\not\!
P\gamma_5 C)_{\alpha\beta}\Lambda_\gamma \nonumber\\
&&+\mathcal{T}_1(P^\nu i\sigma_{\mu\nu}
C)_{\alpha\beta}(\gamma^\mu\gamma_5\Lambda)_\gamma.\label{melement}
\end{eqnarray}
The invariant functions $\mathcal{V}_1,\mathcal{A}_1$ and
$\mathcal{T}_1$ can be represented explicitly:
\begin{equation}
F(a_ip\cdot x)=\int \mathcal{D}\alpha e^{-ip\cdot
x\Sigma_i\alpha_ia_i}F(\alpha_i)\;.
\end{equation}
The integration measure is defined as follows:
\begin{equation}
\int\mathcal{D}\alpha=\int_0^1d\alpha_1d\alpha_2d\alpha_3\delta(\alpha_1+\alpha_2+\alpha_3-1).
\end{equation}

As the usual procedure in QCD sum rule, by taking into account the
dispersion relation and the quark-hadron duality, the hadronic
representation of the correlation function is matched with that
calculated on the light cone. Then after making the Borel
transformation to suppress higher resonance contributions, we get
the following light-cone QCD sum rules:
\begin{eqnarray}
f_\Lambda
F_1(Q^2)e^{-\frac{M_\Lambda^2}{M_B^2}}&=&\{e_u\int_{\alpha_{1_0}}^1d\alpha_1\int_0^{1-\alpha_1}
d\alpha_2\mathcal{A}_1(\alpha_1,\alpha_2,1-\alpha_1-\alpha_2)e^{-\frac{s_1}{M_B^2}}\nonumber\\
&&+e_d\int_{\alpha_{2_0}}^1d\alpha_2\int_0^{1-\alpha_2}d\alpha_1
\mathcal{A}_1(\alpha_1,\alpha_2,1-\alpha_1-\alpha_2)e^{-\frac{s_2}{M_B^2}}\nonumber\\
&&+e_s\int_{\alpha_{3_0}}^1d\alpha_3\int_0^{1-\alpha_3}d\alpha_1\mathcal{A}_1(\alpha_1,1-\alpha_1-\alpha_3,\alpha_3)
e^{-\frac{s_3}{M_B^2}}\},\nonumber\\
F_2(Q^2)&=&0.
\end{eqnarray}
Here $\alpha_{i_0}$ connects with the continuum threshold $s_0$:
\begin{eqnarray}
\alpha_{i_0}&=&\frac{\sqrt{(s_0+Q^2-M_\Lambda^2)^2+4(Q^2+m_i^2)M_\Lambda^2}}{2M_\Lambda^2}-\frac{(s_0+Q^2-M_\Lambda^2)}{2M_\Lambda^2}\label{threshold}
\end{eqnarray}

\subsection{Beyond the leading twist}
It is known that for the nucleon, the leading twist contribution is
rather small, while higher twist contributions are dominant
\cite{Lenz}, which is different from cases of mesons. Therefore, it
is necessary to consider contributions of the higher twist
distribution amplitudes for the calculation of the $\Lambda$ EM form
factors. The usual higher twist contributions come from two
different physical origins. First, as the hard quark propagator will
receive corrections when considering the background gluon field,
there may come contributions from the four-particle (and
five-particle) distribution amplitudes. Second, contributions will
arise from the matrix element of the three-quark operator in Eq.
(\ref{melement}) if we consider other Lorentz structures on the
light cone \cite{Huang,Wang,Braun}. As the first source does not
play a significant role \cite{Lenz,Diehl}, we only consider the
second one in this paper.

In the calculation, only axial-like vector Lorentz structures
contribute, so we merely present the following decomposition of the
matrix element of the three-quark operator:
\begin{eqnarray}
&&4\langle0|\epsilon^{ijk}u_\alpha^i(a_1 x)d_\beta^j(a_2
x)s_\gamma^k(a_3 x)|\Lambda(P)\rangle\nonumber\\
&&=\mathcal{A}_1(\!\not\!P\gamma_5 C)_{\alpha\beta}\Lambda_{\gamma}
+\mathcal{A}_2M(\!\not\!P\gamma_5C)_{\alpha\beta}(\rlap/x\Lambda)_{\gamma}+\mathcal{A}_3M(\gamma_\mu\gamma_5
C)_{\alpha\beta}(\gamma^\mu
\Lambda)_{\gamma}\nonumber\\&&+\mathcal{A}_4M^2(\!\not\!x\gamma_5C)_{\alpha\beta}\Lambda_{\gamma}
+\mathcal{A}_5M^2(\gamma_\mu\gamma_5
C)_{\alpha\beta}(i\sigma^{\mu\nu}x_\nu
\Lambda)_{\gamma}+\mathcal{A}_6M^3(\!\not\!x\gamma_5C)_{\alpha\beta}(\!\not\!x
\Lambda)_{\gamma}.\label{da-def}
\end{eqnarray}
As the above distributions do not have a definite twist (see Ref.
\cite{Lenz,Braun} for a review), the invariant functions
$\mathcal{A}_i$ can be parameterized in terms of distribution
amplitudes $A_1,...\;,A_6$ with a definite twist:
\begin{eqnarray}
&&\mathcal{A}_1=A_1, \nonumber\\
&&2P\cdot x\mathcal{A}_2=-A_1+A_2-A_3, \nonumber\\
&&2\mathcal{A}_3=A_3, \nonumber\\
&&4P\cdot x\mathcal{A}_4=-2A_1-A_3-A_4+2A_5,
\nonumber\\
&&4P\cdot x\mathcal{A}_5=A_3-A_4,\nonumber\\
&&(2P\cdot
x)^2\mathcal{A}_6=A_1-A_2+A_3+A_4-A_5+A_6,\nonumber\\
\label{eqexp}
\end{eqnarray}
where $A_1$ is twist-3, $A_2$ and $A_3$ are twist-4, $A_4$ and $A_5$
are twist-5, and $A_6$ is twist-6 distribution amplitude.

These distribution amplitudes are scale dependent and can be
expanded into contributions of conformal operators. To the leading
order conformal spin accuracy the expansion reads \cite{Wang,Braun}
\begin{eqnarray}
A_1(x_i,\mu)&=&-120x_1x_2x_3\phi_3^0(\mu),\nonumber\\
A_2(x_i,\mu)&=&-24x_1x_2\phi_4^0(\mu),\nonumber\\
A_3(x_i,\mu)&=&-12x_3(1-x_3)\psi_4^0(\mu),\nonumber\\
A_4(x_i,\mu)&=&-3(1-x_3)\psi_5^0(\mu),\nonumber\\
A_5(x_i,\mu)&=&-6x_3\phi_5^0(\mu),\nonumber\\
A_6(x_i,\mu)&=&-2\phi_6^0(\mu).
\end{eqnarray}
With the equation of motion, the six parameters can be expressed in
terms of two independent parameters $f_\Lambda$ and $\lambda_1$. To
the leading order conformal spin accuracy, they are expressed as
\begin{eqnarray}
\phi_3^0&=&\phi_6^0=-f_\Lambda,\nonumber\\
\phi_4^0&=&\phi_5^0=-\frac12(f_\Lambda+\lambda_1),\nonumber\\
\psi_4^0&=&\psi_5^0=\frac12(f_\Lambda-\lambda_1).
\end{eqnarray}

Considering the definition of Eq. (\ref{da-def}) and Eq.
(\ref{eqexp}), the correlation function (\ref{correlator}) is
calculated up to twist $6$ and described explicitly:
\begin{eqnarray}
&&z^\mu T_\mu(P,q)=2e_u(P\cdot z)^2\!\not\! z
\Lambda(P)\int_0^1d\alpha_1\Big\{B_0(\alpha_1)+B_1(\alpha_1)\frac{M_\Lambda^2}{(s_1-(q-P)^2)}\nonumber\\
&&+B_2(\alpha_1)\frac{2M_\Lambda^4}{(s_1-(q-P)^2)^2}\Big\}\frac{1}{s_1-(q-P)^2}+2e_u(P\cdot
z)^2\!\not\! z\!\not\! q\Lambda(P)\nonumber\\
&&\times\int_0^1d\alpha_1\Big\{E_1(\alpha_1)-B_2(\alpha_1)\frac{2M_\Lambda^2}{\alpha_1(s_1-(q-P)^2)^2}\Big\}
\frac{M_\Lambda}{\alpha_1(s_1-(q-P)^2)^2}\nonumber\\
&&+(e_u\rightarrow
e_d,\alpha_1\leftrightarrow\alpha_2,s_1\rightarrow
s_2,B_i\rightarrow C_i,E_1\rightarrow E_2)\nonumber\\
&&+(e_u\rightarrow
e_s,\alpha_1\rightarrow\alpha_3,\alpha_2\rightarrow\alpha_1,s_1\rightarrow
s_3,B_i\rightarrow D_i,E_1\rightarrow E_3).
\end{eqnarray}
Here the following notation is used for convenience:
\begin{eqnarray}
B_0(\alpha_1)&=&\int_0^{1-\alpha_1}d\alpha_2A_1(\alpha_1,\alpha_2,1-\alpha_1-\alpha_2),\nonumber\\
B_1(\alpha_1)&=&2\widetilde A_1(\alpha_1)-\widetilde
A_2(\alpha_1)+\widetilde A_3(\alpha_1)+\widetilde
A_4(\alpha_1)-\widetilde
A_5(\alpha_1),\nonumber\\
B_2(\alpha_1)&=& \widetilde{\widetilde
A}_1(\alpha_1)-\widetilde{\widetilde
A}_2(\alpha_1)+\widetilde{\widetilde
A}_3(\alpha_1)+\widetilde{\widetilde A}_4(\alpha_1)
-\widetilde{\widetilde A}_5(\alpha_1)+\widetilde{\widetilde A}_6(\alpha_1),\nonumber\\
C_0(\alpha_2)&=&\int_0^{1-\alpha_2}d\alpha_1A_1(\alpha_1,\alpha_2,1-\alpha_1-\alpha_2),\nonumber\\
C_1(\alpha_2)&=&2\widetilde A_1(\alpha_2)-\widetilde
A_2(\alpha_2)+\widetilde A_3(\alpha_2)+\widetilde
A_4(\alpha_2)-\widetilde
A_5(\alpha_2),\nonumber\\
C_2(\alpha_2)&=&\widetilde{\widetilde
A}_1(\alpha_2)-\widetilde{\widetilde
A}_2(\alpha_2)+\widetilde{\widetilde
A}_3(\alpha_2)+\widetilde{\widetilde A}_4(\alpha_2)
-\widetilde{\widetilde A}_5(\alpha_2)+\widetilde{\widetilde
A}_6(\alpha_2),\nonumber\\
D_0(\alpha_3)&=&\int_0^{1-\alpha_3}d\alpha_1A_1(\alpha_1,1-\alpha_1-\alpha_3,\alpha_3),\nonumber\\
D_1(\alpha_3)&=&2\widetilde A_1(\alpha_3)-\widetilde
A_2(\alpha_3)+\widetilde A_3(\alpha_3)+\widetilde
A_4(\alpha_3)-\widetilde
A_5(\alpha_3),\nonumber\\
D_2(\alpha_3)&=& \widetilde{\widetilde
A}_1(\alpha_3)-\widetilde{\widetilde
A}_2(\alpha_3)+\widetilde{\widetilde
A}_3(\alpha_3)+\widetilde{\widetilde A}_4(\alpha_3)
-\widetilde{\widetilde A}_5(\alpha_3)+\widetilde{\widetilde
A}_6(\alpha_3),\nonumber\\
E_1(\alpha_1)&=&-\widetilde A_1(\alpha_1)+\widetilde
A_2(\alpha_1)-\widetilde A_3(\alpha_1),\nonumber\\
E_2(\alpha_2)&=&-\widetilde A_1(\alpha_2)+\widetilde
A_2(\alpha_2)-\widetilde A_3(\alpha_2),\nonumber\\
E_3(\alpha_3)&=&-\widetilde A_1(\alpha_3)+\widetilde
A_2(\alpha_3)-\widetilde A_3(\alpha_3),
\end{eqnarray}
where the distribution amplitudes with a ``tilde" are defined as
\begin{eqnarray}
\widetilde
A(\alpha_1)&=&\int_0^{\alpha_1}d\alpha_1'\int_0^{1-\alpha_1'}d\alpha_2A(\alpha_1',\alpha_2,1-\alpha_1'-\alpha_2),\nonumber\\
\widetilde
A(\alpha_2)&=&\int_0^{\alpha_2}d\alpha_2'\int_0^{1-\alpha_2'}d\alpha_1A(\alpha_1,\alpha_2',1-\alpha_1-\alpha_2'),\nonumber\\
\widetilde
A(\alpha_3)&=&\int_0^{\alpha_3}d\alpha_3'\int_0^{1-\alpha_3'}d\alpha_1A(\alpha_1,1-\alpha_1-\alpha_3',\alpha_3'),\nonumber\\
\widetilde{\widetilde
A}(\alpha_1)&=&\int_0^{\alpha_1}d\alpha_1'\int_0^{\alpha_1'}d\alpha_1''\int_0^{1-\alpha_1''}d\alpha_2A(\alpha_1'',\alpha_2,1-\alpha_1''-\alpha_2),\nonumber\\
\widetilde{\widetilde
A}(\alpha_2)&=&\int_0^{\alpha_2}d\alpha_2'\int_0^{\alpha_2'}d\alpha_2''\int_0^{1-\alpha_2''}d\alpha_1 A(\alpha_1,\alpha_2'',1-\alpha_1-\alpha_2''),\nonumber\\
\widetilde{\widetilde
A}(\alpha_3)&=&\int_0^{\alpha_3}d\alpha_3'\int_0^{\alpha_3'}d\alpha_3''\int_0^{1-\alpha_3''}d\alpha_1
A(\alpha_1,1-\alpha_1-\alpha_3'',\alpha_3'').
\end{eqnarray}
These results stem from the partial integration in $\alpha_1,
\alpha_2$ and $\alpha_3$, respectively, which is employed to
eliminate the factors $1/(P\cdot x)^n$ in the calculation.

Now by using the Borel transformation and subtraction similarly to
that in Ref. \cite{Lenz}, we arrive at the final sum rules:
\begin{eqnarray}
f_\Lambda
F_1(Q^2)e^{-\frac{M_\Lambda^2}{M_B^2}}&=&e_u\Big\{\int_{\alpha_{1_0}}^1d\alpha_1\{B_0(\alpha_1)
+B_1(\alpha_1)\frac{M_\Lambda^2}{M_B^2}+B_2(\alpha_1)\frac{M_\Lambda^4}{M_B^4}\}
e^{-\frac{s_1}{M_B^2}}+\{B_1(\alpha_{1_0})\nonumber\\
&&+B_2(\alpha_{1_0})\frac{M_\Lambda^2}{M_B^2}-\frac{d}{d\alpha_{1_0}}
\frac{\alpha_{1_0}^2M_\Lambda^2B_2(\alpha_{1_0})}{Q^2+\alpha_{1_0}^2M_\Lambda^2}\}
\frac{M_\Lambda^2\alpha_{1_0}^2}{Q^2+\alpha_{1_0}^2M_\Lambda^2}e^{-\frac{s_0}{M_B^2}}\Big\}\nonumber\\
&&+e_d\Big\{\int_{\alpha_{2_0}}^1d\alpha_2\{C_0(\alpha_2)
+C_1(\alpha_2)\frac{M_\Lambda^2}{M_B^2}+
C_2(\alpha_2)\frac{M_\Lambda^4}{M_B^4}\}e^{-\frac{s_2}{M_B^2}}\nonumber\\
&&+\{C_1(\alpha_{2_0})
+C_2(\alpha_{2_0})\frac{M_\Lambda^2}{M_B^2}-\frac{d}{d\alpha_{2_0}}
\frac{\alpha_{2_0}^2M_\Lambda^2C_2(\alpha_{2_0})}{Q^2+\alpha_{2_0}^2M_\Lambda^2}\}
\frac{M_\Lambda^2\alpha_{2_0}^2}{Q^2+\alpha_{2_0}^2M_\Lambda^2}e^{-\frac{s_0}{M_B^2}}\Big\}\nonumber\\
&&+e_s\Big\{\int_{\alpha_{3_0}}^1d\alpha_3\{D_0(\alpha_3)
+D_1(\alpha_3)\frac{M_\Lambda^2}{M_B^2}+D_2(\alpha_3)\frac{M_\Lambda^4}{M_B^4}\}
e^{-\frac{s_3}{M_B^2}}+\{D_1(\alpha_{3_0})\nonumber\\
&&+D_2(\alpha_{3_0})\frac{M_\Lambda^2}{M_B^2}-\frac{d}{d\alpha_{3_0}}
\frac{\alpha_{3_0}^2M_\Lambda^2D_2(\alpha_{3_0})}{Q^2+m_s^2+\alpha_{3_0}^2M_\Lambda^2}\}
\frac{M_\Lambda^2\alpha_{3_0}^2}{Q^2+m_s^2+\alpha_{3_0}^2M_\Lambda^2}e^{-\frac{s_0}{M_B^2}}\Big\},\nonumber\\
\label{sumrule1}
\end{eqnarray}
and
\begin{eqnarray}
f_\Lambda
F_2(Q^2)e^{-\frac{M_\Lambda^2}{M_B^2}}&=&2\Big\{e_u\big\{\int_{\alpha_{1_0}}^1d\alpha_1
\{\frac{M_\Lambda^2}{\alpha_1M_B^2}E_1(\alpha_1)-\frac{M_\Lambda^4}{\alpha_1M_B^4}B_2(\alpha_1)\}e^{-\frac{s_1}{M_B^2}}\nonumber\\
&&+\{\frac{1}{\alpha_{1_0}}E_1(\alpha_{1_0})
-\frac{M_\Lambda^2}{\alpha_{1_0}M_B^2}B_2(\alpha_{1_0})+\frac{d}{d\alpha_{1_0}}\frac{\alpha_{1_0}M_\Lambda^2}
{Q^2+\alpha_{1_0}^2M_\Lambda^2}B_2(\alpha_{1_0})\}\nonumber\\
&&\times \frac{\alpha_{1_0}^2M_\Lambda^2}{Q^2+\alpha_{1_0}^2M_\Lambda^2}e^{-\frac{s_0}{M_B^2}}\big\}\nonumber\\
&&+e_d\big\{\int_{\alpha_{2_0}}^1d\alpha_2\{\frac{M_\Lambda^2}{\alpha_2M_B^2}E_2(\alpha_2)
-\frac{M_\Lambda^4}{\alpha_2M_B^4}C_2(\alpha_2)\} e^{-\frac{s_2}{M_B^2}}\nonumber\\
&&+\{\frac{1}{\alpha_{2_0}}E_2(\alpha_{2_0})-\frac{M_\Lambda^2}{\alpha_{2_0}M_B^2}C_2(\alpha_{2_0})\nonumber\\
&&+\frac{d}{d\alpha_{2_0}}\frac{\alpha_{2_0}M_\Lambda^2}
{Q^2+\alpha_{2_0}^2M_\Lambda^2}C_2(\alpha_{2_0})\}\frac{\alpha_{2_0}^2M_\Lambda^2}{Q^2+\alpha_{2_0}^2M_\Lambda^2}e^{-\frac{s_0}{M_B^2}}\big\}\nonumber\\
&&+e_s\big\{\int_{\alpha_{3_0}}^1d\alpha_3\{\frac{M_\Lambda^2}{\alpha_3M_B^2}E_3(\alpha_3)
-\frac{M_\Lambda^4}{\alpha_3M_B^4}D_2(\alpha_3)\}
e^{-\frac{s_3}{M_B^2}}\nonumber\\
&&+\{\frac{1}{\alpha_{3_0}}E_3(\alpha_{3_0})
-\frac{M_\Lambda^2}{\alpha_{3_0}M_B^2}D_2(\alpha_{3_0})+\frac{d}{d\alpha_{3_0}}\frac{\alpha_{3_0}M_\Lambda^2}
{Q^2+m_s^2+\alpha_{3_0}^2M_\Lambda^2}\nonumber\\
&&\times D_2(\alpha_{3_0})\}
\frac{\alpha_{3_0}^2M_\Lambda^2}{Q^2+m_s^2+\alpha_{3_0}^2M_\Lambda^2}e^{-\frac{s_0}{M_B^2}}\big\}\Big\}.
\label{sumrule2}
\end{eqnarray}

\section{Numerical analysis} \label{sec3}
\subsection{Determination of the parameters $f_\Lambda$ and $\lambda_1$}
The parameters $f_\Lambda$ and $\lambda_1$ can be determined by the
QCD sum rule method. The sum rules begin with the following
correlation functions:
\begin{equation}
\Pi(q^2)=i\int d^4xe^{iq\cdot x}\langle0|T\{j_i(x)\bar
j_i(0)\}|0\rangle,
\end{equation}
where the interpolating currents are
\begin{eqnarray}
j_1(x)&=&\epsilon_{ijk}(u^i(0)C\gamma_5\!\not\!
zd^j(0))\!\not\! zs^k(0),\nonumber\\
j_2(x)&=&\epsilon_{ijk}(u^i(x)C\gamma_5\gamma_\mu d^j(x))\gamma_\mu
s^k(x),
\end{eqnarray}
and following the standard QCD sum rule procedure, the results are
given by
\begin{eqnarray}
(4\pi)^4f_\Lambda^2e^{-\frac{M_\Lambda^2}{M_B^2}}&=&\frac25\int_{m_s^2}^{s_0}s(1-x)^5e^{-\frac{s}{M_B^2}}ds
-\frac b3\int_{m_s^2}^{s_0}x(1-x)(1-2x)e^{-\frac{s}{M_B^2}}\frac{ds}{s},\nonumber\\
4(2\pi)^4\lambda_1^2M_\Lambda^2e^{-\frac{M_\Lambda^2}{M_B^2}}&=&\int_{m_s^2}^{s_0}\frac{s^2}{2}[(1-x^2)(1-8x+x^2)
-12x^2\ln(x)]e^{-\frac{s}{M_B^2}}ds\nonumber\\
&&+\frac{b}{12}\int_{m_s^2}^{s_0}(1-x)^2e^{-\frac{s}{M_B^2}}ds-\frac43a^2e^{-\frac{m_s^2}{M_B^2}}+m_sa_s\int_{m_s^2}^{s_0}e^{\frac{s}{M_B^2}}ds,
\end{eqnarray}
where $x=m_s^2/s$.

The sum rules reveal that they can only give the absolute values of
the two parameters. The relative sign of $f_\Lambda$ and $\lambda_1$
can be determined by the following correlation function:
\begin{equation}
\Pi(q^2)=i\int d^4xe^{iq\cdot x}\langle0|T\{j_1(x)\bar
j_2(0)\}|0\rangle,
\end{equation}
from which the sum rule of $f_\Lambda\lambda_1^*$ is given as
follows:
\begin{eqnarray}
4(2\pi)^4f_\Lambda\lambda_1^*M_\Lambda e^{-\frac{M_\Lambda^2}{M_B^2}}&=&\int_{m_s^2}^{s_0}\frac{m_s}{6}s[(1-x)(3+13x-5x^2+x^3)
+12x\ln(x)]e^{-\frac{s}{M_B^2}}ds \nonumber\\
&&+\frac{b}{12}\int_{m_s^2}^{s_0}m_s(1-x)[1-\frac13(1-x)(5-\frac2x)]e^{-\frac{s}{M_B^2}}\frac{ds}{s}\nonumber\\
&&+\frac{a_s}{24}\int_{m_s^2}^{s_0}e^{-\frac{s}{M_B^2}}ds.
\label{relasign}
\end{eqnarray}

In the numerical analysis we adopt the standard values:
\begin{eqnarray}
&&a=-(2\pi)^2\langle \bar qq\rangle=0.55\;\mbox{GeV}^3,\nonumber\\
&&b=(2\pi)^2 \langle \alpha_s G^2/ \pi \rangle=0.47\;\mbox{GeV}^4,\nonumber\\
&&a_s=m_0^2a,\hspace{0.4cm} m_0^2=0.8\;\mbox{GeV}^2.
\end{eqnarray}
The threshold is set to be $s_0=1.6^2\;\mbox{GeV}^2$, and the
working window for the Borel parameter is $1\;\mbox{GeV}^2<M_B^2<2\;
\mbox{GeV}^2$. The sum rule (\ref{relasign}) shows that
$f_\Lambda\lambda_1^*$ is positive. Here $f_\Lambda$ is taken to be
positive and the parameters have the following numerical values:
\begin{eqnarray}
f_\Lambda&=&(5.9\pm0.2)\times10^{-3}\;\mbox{GeV}^2,\nonumber\\
\lambda_1&=&(1.0\pm0.3)\times10^{-2}\;\mbox{GeV}^2.\label{nonpara}
\end{eqnarray}
\subsection{Analysis of the sum rules}
\begin{figure}
\begin{minipage}{7cm}
\epsfxsize=7cm \centerline{\epsffile{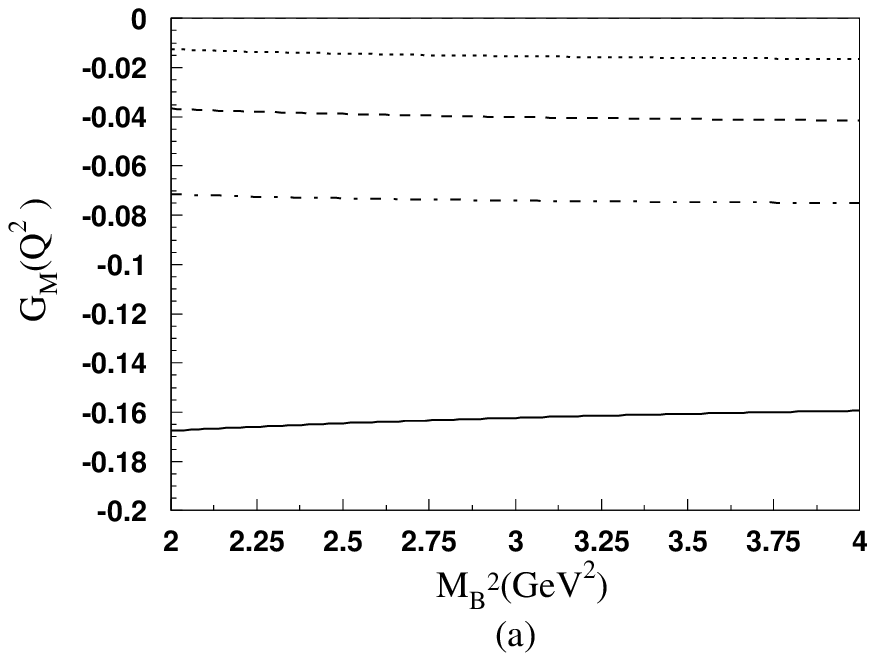}}
\end{minipage}
\hfill
\begin{minipage}{7cm}
\epsfxsize=7cm \centerline{\epsffile{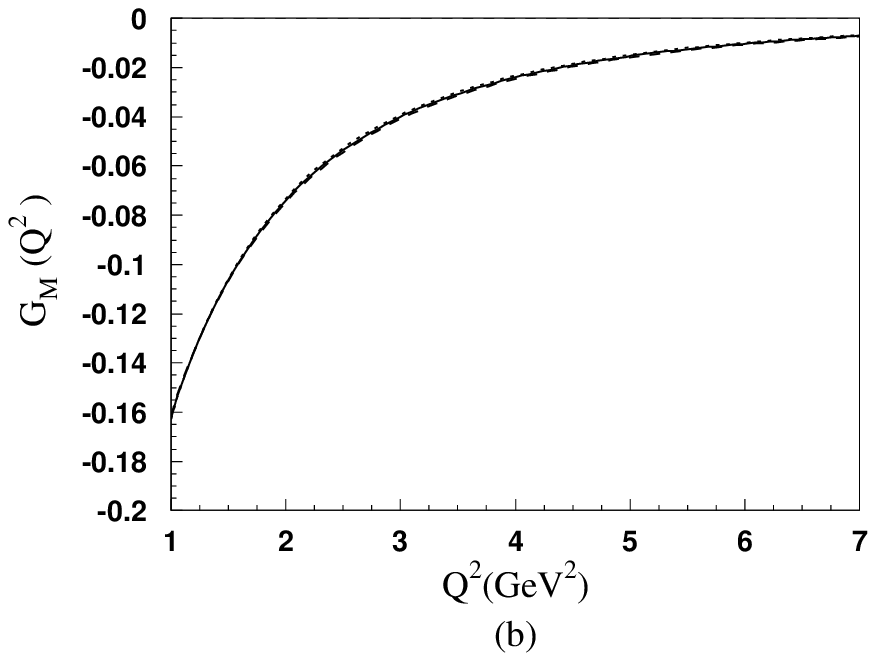}}
\end{minipage}
\caption{}\label{fig1}
\end{figure}

Before the numerical analysis, we firstly specify the input
parameters used in the light-cone sum rules. The mass of the
$\Lambda$ baryon is given by the Particle Data Group (PDG)
\cite{PDG}: $M_\Lambda=1.116\;\mbox{GeV}$. The mass of the strange
quark is chosen to be $m_s=0.15\;\mbox{GeV}$, and the continuum
threshold $s_0=2.45-2.65\;\mbox{GeV}^{2}$. For the auxiliary Borel
parameter $M_B^2$, a working window in which the results vary mildly
is required. The choice of the Borel parameter should satisfy two
conditions simultaneously. On the one hand, as the higher twist
contributions are proportional to terms $(1/M_B^2)^n$ ($n=1,2,...$),
$M_B^2$ should be large enough to suppress the higher twist
contributions. On the other hand, $M_B^2$ cannot be too large in
case that higher resonance and continuum contributions become
dominant. In the calculation, the Borel parameter varies in the
range $2\;\mbox{GeV}^{2}\leq M_B^2\leq 4\;\mbox{GeV}^{2}$. Fig.
\ref{fig1}(a) gives the dependence of the magnetic form factors on
the Borel parameter at different point of $Q^2$. One can see from
the figure that the results are almost independent of the Borel
parameter.

\begin{figure}
\begin{minipage}{7cm}
\epsfxsize=7cm \centerline{\epsffile{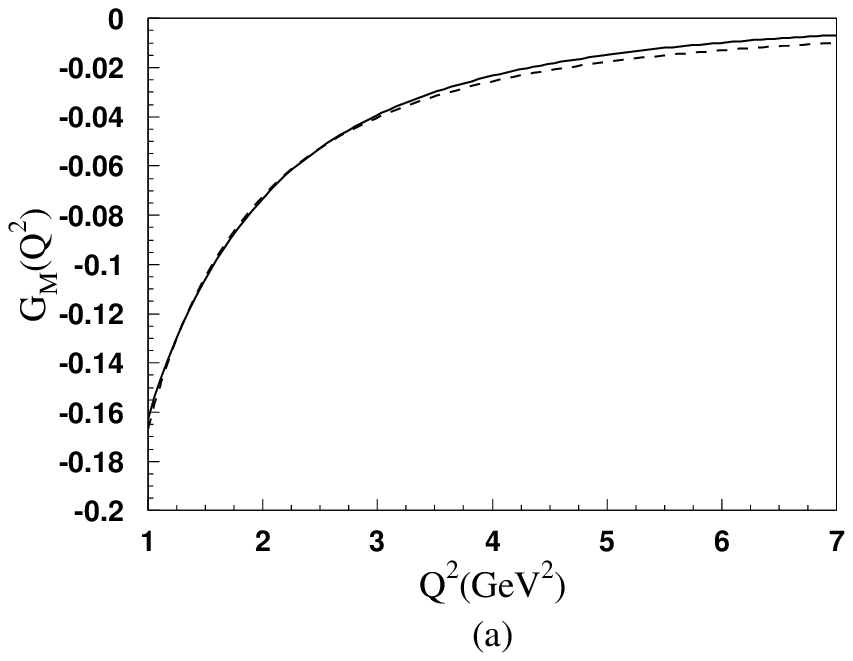}}
\end{minipage}
\hfill
\begin{minipage}{7cm}
\epsfxsize=7cm \centerline{\epsffile{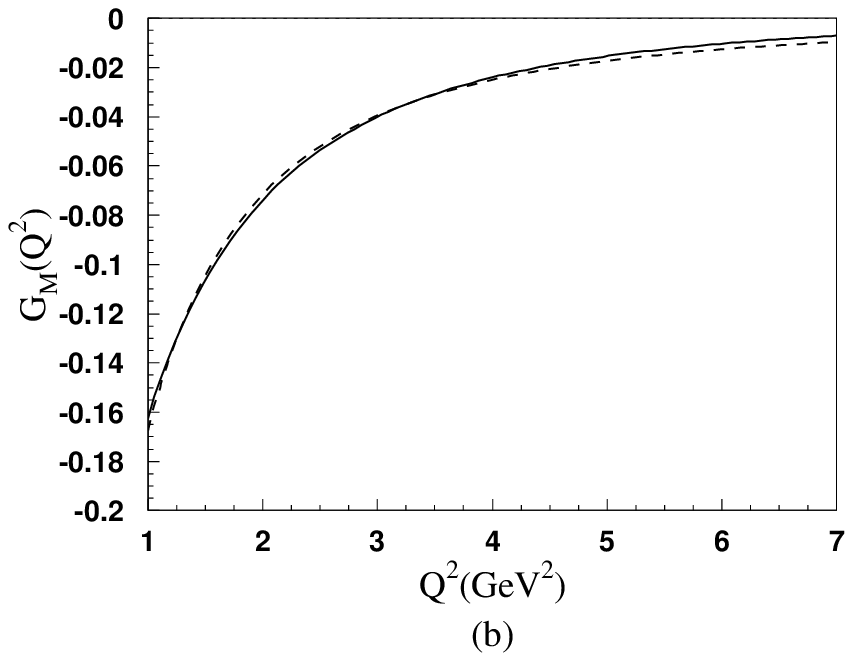}}
\end{minipage}
\caption{}\label{fig2}
\end{figure}

\begin{figure}
\begin{minipage}{7cm}
\epsfxsize=7cm \centerline{\epsffile{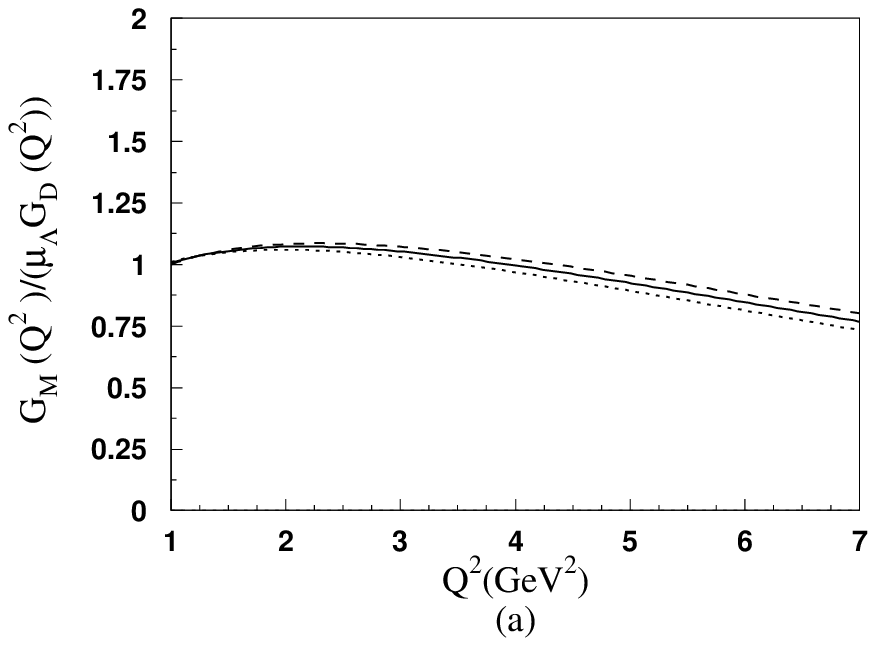}}
\end{minipage}
\hfill
\begin{minipage}{7cm}
\epsfxsize=7cm \centerline{\epsffile{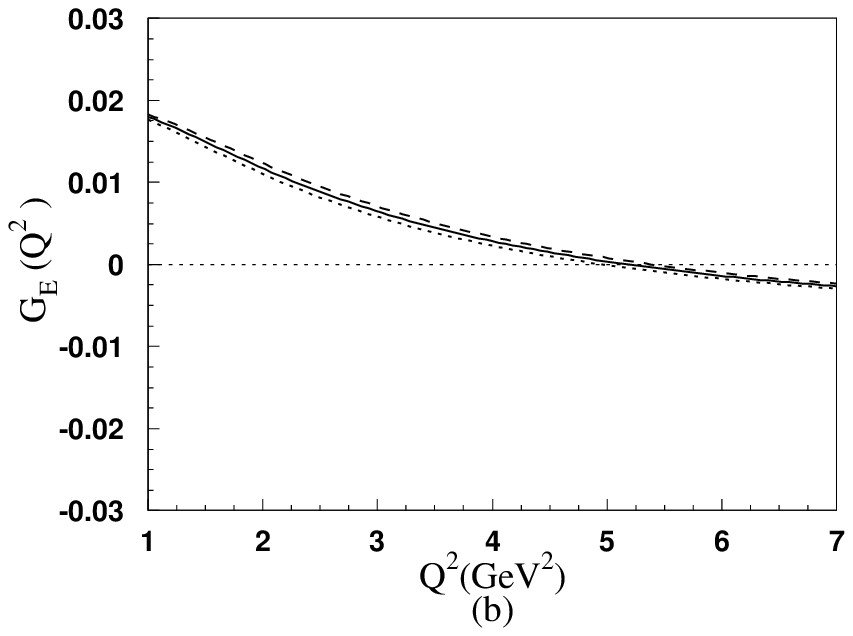}}
\end{minipage}
\caption{}\label{fig3}
\end{figure}

In the following numerical analysis of the sum rules the Borel
parameter is taken to be $M_B^2=3\;\mbox{GeV}^{2}$, and the
nonperturbative parameters $f_\Lambda$ and $\lambda_1$ are the
central values in Eq. (\ref{nonpara}). Fig. \ref{fig1}(b) gives the
dependence of the magnetic form factor $G_M(Q^2)$ on the momentum
transfer $Q^2$. The figure shows that the $Q^2$-dependence of the
magnetic form factor $G_M(Q^2)$ is in accordance with our assumption
in Eq. (\ref{dipole}). In order to estimate the magnetic moment, the
magnetic form factor is fitted by the dipole formula (\ref{dipole}).
The fits are shown in Fig. \ref{fig2}, from which we evaluate the
numerical values of the two parameters simultaneously:
$\mu_\Lambda=-(0.64\pm0.04)\mu_N$ and
$m_0^2=(0.89\pm0.04)\;\mbox{GeV}^{2}$. In Fig. \ref{fig3}(a) we also
display the $Q^2$-dependence of the value $ G_M/(\mu_\Lambda G_D)$.
The parameter $m_0^2$ used in the dipole formula (\ref{dipole}) is
the central value obtained above, while the magnetic moment
$\mu_\Lambda=0.613\mu_N$ comes from Ref. \cite{PDG}. It is shown
from the figure that the result deviates from $1$ when $Q^2$ becomes
large, which is due to the fact that the absolute value of
$G_M(Q^2)$ form factor decreases with $Q^2$, and the influence of
the threshold $s_0$ becomes important at large momentum transfer.

Tab. \ref{tabel} lists results of $\Lambda$ magnetic moment from
various approaches. In comparison with results in Tab. \ref{tabel},
our estimation is in accordance with data provided by PDG and other
theoretical predictions.

Fig. \ref{fig3}(b) is the $Q^2$-dependence of the electric form
factor $G_E(Q^2)$. The figure shows that the electric form factor
changes sign at a finite value of $Q^2$, which means that in the
large momentum transfer the massive $s$ quark play a more important
role in determining the electric density of the baryon. This is
different from that of the neutron \cite{Lenz}. A similar conclusion
has been given in Ref. \cite{Cauteren}. The difference to our
calculation is that the electric form factor is negative at lower
$Q^2$ in their calculation, while another result from Ref.
\cite{Kim} within the frame-work of the chiral quark/soliton model
showed that at small $Q^2$ this form factor is positive, which is in
contradiction to the result from Ref. \cite{Cauteren}.

\begin{figure}
\begin{minipage}{7cm}
\epsfxsize=7cm \centerline{\epsffile{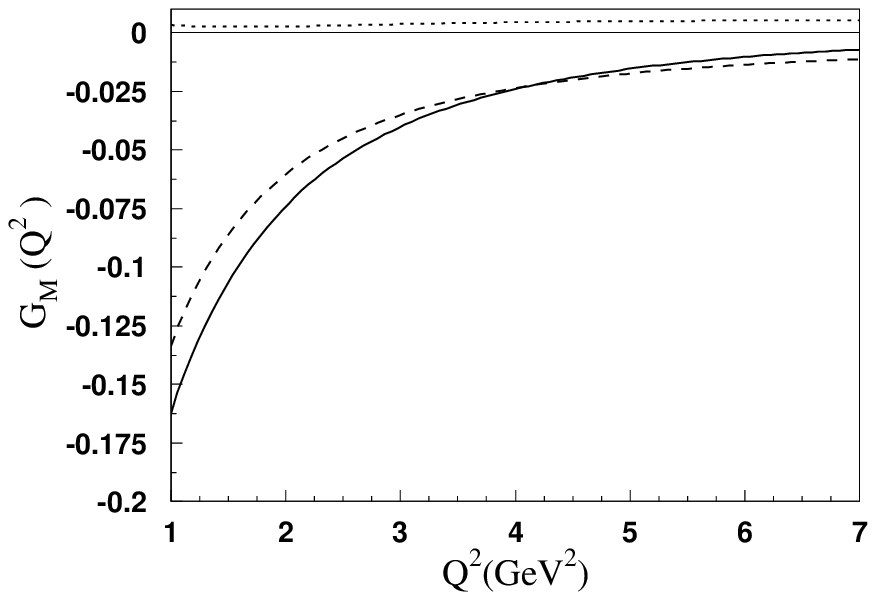}}
\end{minipage}
\caption{}\label{fig4}
\end{figure}

Finally, contributions of distribution amplitudes with different
twist are calculated for $G_M(Q^2)$, which is shown in Fig.
\ref{fig4}. It can be concluded that it is twist $4$ but not leading
twist contributions that dominate the form factor. This stems from
the structure of the leading order distribution amplitude of the
$\Lambda$ baryon, which is symmetric on the $d$ and $s$ quark if the
mass of $s$ quark approaches zero. Contributions of $d$ and $s$
quark can be canceled by the contribution of $u$ quark in the
approach. The results are expected to be better if more information
about the distribution amplitudes of the $\Lambda$ baryon is known.

To summarize, we provide a fit approach to predict the magnetic
moment of a hadron. The $Q^2$-dependence EM form factors of the
$\Lambda$ baryon are calculated in the framework of the light-cone
sum rule up to twist $6$. The magnetic form factor is fitted by the
dipole formula to estimate the magnetic moment of the baryon. Our
estimation is in accordance with the existing results. As there lack
experimental data on the baryon, we only give theoretical
investigations on the form factors as what have been done on
nucleons. Analysis on the electric form factor shows that it changes
sign at a finite large momentum transfer, which is expected to be
tested by the future experiments. Studies on contributions of the
distribution amplitudes with different twist show that it is the
twist $4$ but not the leading twist contributions that dominate the
result.

\acknowledgments This work was supported in part by the National
Natural Science Foundation of China under Contract No.10675167.

\newpage

{\bf{Figure and table captions}}

\begin{center}
\begin{minipage}{14cm}
{\sf Fig. \ref{fig1}.}{(a) Dependence of the $\Lambda$ magnetic form
factor $G_M(Q^2)$ on the Borel parameter. The lines correspond to
the points $Q^2=1,\;2,\;3,\;5\;\mbox{GeV}^{2}$ from the bottom up
with the threshold $s_0=2.55\;\mbox{GeV}^2$. (b) $Q^2$-dependence of
the magnetic form factor $G_M$. The lines correspond to the
threshold $s_0=2.45-2.65\;\mbox{GeV}^2$ from up down.}
\end{minipage}
\end{center}

\begin{center}
\begin{minipage}{14cm}
{\sf Fig. \ref{fig2}.}{ Fittings of the form factor $G_M(Q^2)$ by
$\mu_\Lambda/(1+Q^2/m_0^2)^2$ where the dashed lines are the fits.
Figures $(a),\;(b)$ correspond to threshold $s_0=2.45,\;
2.65\;\mbox{GeV}^{2}$, respectively.}
\end{minipage}
\end{center}

\begin{center}
\begin{minipage}{14cm}
{\sf Fig. \ref{fig3}.}{ The $Q^2$-dependence of the form factor
$G_M/(\mu_\Lambda G_D)$ (a) and $G_E$ (b). The lines correspond the
threshold $s_0=2.45,\;2.55,\;2.65\;\mbox{GeV}^{2}$ from up down.}
\end{minipage}
\end{center}

\begin{center}
\begin{minipage}{14cm}
{\sf Fig. \ref{fig4}.}{ Contributions of different twist for
$G_M(Q^2)$ at the threshold $s_0=2.55\;\mbox{GeV}^2$. The dotted
line, dashed line and the solid line correspond to twist $3$, twist
$4$ and all contributions, respectively.}
\end{minipage}
\end{center}

\begin{center}
\begin{minipage}{14cm}
{\sf Tab. \ref{tabel}.}{ The magnetic moment of the $\Lambda$ baryon
from various models.}
\end{minipage}
\end{center}

\clearpage
\newpage

\clearpage
\newpage

\begin{table}[h]
\caption{}
\begin{center}
\begin{tabular}{|c|c|c|c|c|c|c|c|c|c|c|}
\hline Model&PDG&QCDSR&SQM&QCDSA &$\chi$PT &SKRM&NQM
&EQLA&GSE&LCSR \\
&\cite{PDG}&\cite{Pasupathy}&\cite{Franklin}&\cite{Kerbikov}&\cite{Puglia}&\cite{Park}&
\cite{Contreras}&\cite{Simonov}&\cite{Bartelski}&\cite{Aliev}\\
 \hline
 $\mu_\Lambda(\mu_N)$ & $-0.613$ & $-0.50/$ & $-0.67$\hspace{0.1cm} &
$-0.69$ & $-0.613\hspace{0.1cm}$& $-0.60$ & $-0.63$ & $-0.60$ & $-0.606\hspace{0.1cm} $ & $-0.7$\\
&&$-0.54$&&&&&&&&\\
\hline
\end{tabular}
\end{center}\label{tabel}
\end{table}

\end{document}